\documentclass[12pt,a4paper]{article}
\usepackage{amsfonts,latexsym}
\usepackage{graphicx,color}
\usepackage{dcolumn}
\usepackage{graphicx}
\usepackage{amsmath}
\usepackage{amsfonts}
\usepackage{amssymb}

\renewcommand{\thefootnote}{\fnsymbol{footnote}}

\begin{document}

\vspace{12mm}

\begin{center}
{{{\Large {\bf Superradiant instability of a massive scalar \\ on the  Kerr spacetime  }}}}\\[10mm]

Yun Soo Myung$^{a,b}$\footnote{e-mail address: ysmyung@inje.ac.kr}\\[8mm]

{${}^a$Institute of Basic Sciences and Department  of Computer Simulation,  Inje University Gimhae 50834, Korea\\[0pt]}

{${}^b$Asia Pacific Center for Theoretical Physics, Pohang 37673, Korea}

\end{center}
\vspace{2mm}

\begin{abstract}
We investigate  the superradiant instability of Kerr  black holes under a  massive scalar perturbation.
We obtain a potential  $V_i(r)$ when expanding the scalar potential $V_K(r)$ for large $r$.
The Newton potential $V_1(r)$ and the far-region potential $V_2(r)$ are used to explore the superradiant instability, while $V_3(r)$ matches a geodesic potential $V_{gK}(r)$
for a neutral particle moving around the Kerr black hole. Thus, $V_{gK}(r)$ is employed to fix the separation constant.
We obtain  a  condition  for a trapping well to possess a quasibound state in  the  Kerr black holes by analyzing the Newton potential $V_1(r)$and far-region wave functions obtained from $V_2(r)$.
The other condition for no trapping well (a tiny well) is found to generate an asymptotic  bound state. Finally, we discuss an ultralight boson whose potential has a tiny well located at asymptotic region.
\end{abstract}
\vspace{5mm}

\vspace{1.5cm}

\hspace{11.5cm}
\newpage
\renewcommand{\thefootnote}{\arabic{footnote}}
\setcounter{footnote}{0}


\section{Introduction}
A merging of two black holes in vacuum is a meticulous prediction of general relativity, which has confirmed recently by gravitational wave observation of the LIGO/Virgo Collaborartion~\cite{LIGOScientific:2016aoc,LIGOScientific:2021sio}. If dark matter clusters surround black hole binary, it would affect  the inspiral through dynamical
friction~\cite{Kavanagh:2020cfn,Coogan:2021uqv}. Ultralight bosons (axions) arisen in the string landscape~\cite{Arvanitaki:2009fg}  may be dark matter candidates~\cite{Hui:2016ltb}. On the other hand, superradiant instability of rotating black hole  could be a continuous source of gravitational waves. That is, untralight bosons  could trigger superradiant instability of rotating black holes. This superradiant instability produces a G-atom (gravitational atom like hydrogen atom) which consists of the central black hole and a surrounding axion cloud in quantum states. Since the axion cloud of a G-atom is not spherically symmetric and rotating, it emits gravitational waves~\cite{Kodama:2011zc}. These gravitational  waves could be probed  with future gravitational wave observations.

If a massive scalar is impinging upon a Kerr black hole, its mass $\mu$ would act as a reflecting mirror and lead to superradiant instability when the parameters of the black hole and scalar are in certain parameter space~\cite{Press:1972zz,Damour:1976kh,Brito:2015oca}. Superradiant instability depends on two parameters of $a=J/M$  and $M\mu$ where $J$ is an angular momentum of the black hole and $M$  is a mass of the black hole. The superradiant instability gets stronger as $a$ and $\mu$ increase~\cite{Cardoso:2004nk}. The efficiency of the superradiant process depends on the ratio of the black hole mass to the Compton wavelength of the scalar (gravitational fine structure constant): $r_g/\lambda_c=M\mu$.
Superradiant instability of the Kerr black hole was found  for $M\mu\gg1$~\cite{Zouros:1979iw}, $M\mu\ll1$~\cite{Detweiler:1980uk}, and   $M\mu\le0.5$~\cite{Dolan:2007mj}.
Noting that the ratio $M\mu$ determines the size of cloud~\cite{Ghosh:2021uqw}, a non-relativistic cloud ($M\mu \ll1$) would be quite far away from the black hole. For $M\mu\leq1$, the cloud  would be close to the black hole. For $M\mu \sim (0.01,0.1)$, the cloud grows quickly and is long lived on astrophysical timescales~\cite{Baumann:2022pkl}.

It is worth mentioning that  the presence of a trapping well is a key to achieve the superradiant instability since scalar modes could be localized in the trapping well (local minimum),   amplified to form a quasibound state by superradiance, and triggered an instability.
A typical scalar potential for the instability  has  a shape of ergo-barrier-well-mirror which   is responsible for generating   quasibound states  if two conditions of  $\omega<\omega_c$ and $\omega<\mu$ are satisfied~\cite{Arvanitaki:2010sy,Konoplya:2011qq}. The former (latter) condition represents  the superradiance (bound state) condition. If  a tiny well (no trapping well) exits under a massive scalar wave propagation, the black hole may be  superradiantly stable.

Before we proceed, we note  that  a shortened form of the potential
was employed to sketch the superradiant instability for rotating black branes and strings~\cite{Cardoso:2005vk} because  $\Psi_{\ell m}=rR_{\ell m}$ and a modified tortoise coordinate $z_*=\int (r^2/\Delta)dr$ are used as in Ref.~\cite{Furuhashi:2004jk}. Also, we wish to point out that a condition for a trapping well was derived  as $\mu/\sqrt{2}<\omega<\mu$ by making use of an inappropriate potential $V(r)=\omega^2-(U(r)+M^2-a^2)/\Delta^2$ and its asymptotic form $V_{\rm asy}(r)=\mu^2+(2M\mu^2-4M\omega^2)/r$~\cite{Hod:2012zza}. The same potential $V_{\rm asy}(r)$ was used to derive two conditions for no trapping well as $\omega<\mu/\sqrt{2}$ and $r_-/r_+<0.802$~\cite{Huang:2019xbu}. Recently,  the superradiant stability of braneworld extremal Kerr and Kerr-Newman black holes was investigated  under a
massive scalar perturbation by choosing this type of potential~\cite{Biswas:2021gvq}.

In this work, we wish to study the superradiant instability of the Kerr black holes under a massive scalar perturbation  by analyzing its asymptotic potential and far-region wave function.  First of all, we derive a potential  $V_i(r)$ for up to $(1/r)^i$ with $i=1,\cdots,8$  when expanding the scalar potential $V_{\rm K}(r)$ for large $r$.
The asymptotic (Newton) potential $V_1(r)$ and the far-region potential $V_2(r)$ are used to investigate the superradiant instability, while $V_3(r)$ matches a geodesic potential $V_{\rm gK}(r)$
for a neutral particle moving around the Kerr black hole. Hence, $V_{\rm gK}(r)$ is employed to fix the separation constant as $K_{lm}=l(l+1)+a^2(\mu^2-\omega^2)$.
We obtain  a  condition  for a trapping well to possess a quasibound state in  the  Kerr black holes by analyzing the Newton potential $V_1(r)$  and far-region wave functions obtained from $V_2(r)$.
The other condition for no trapping well (a tiny well) is found to generate an asymptotic  bound state. This work would be  a newly analysis for superradiant instability of Kerr black holes under a massive scalar perturbation. Finally, we discuss an ultralight boson with a tiny well located at asymptotic region.

\section{A massive scalar in  Kerr black holes }
We introduce  the  Kerr
black hole written by Boyer-Lindquist coordinates
\begin{eqnarray}
ds^2_{\rm K} &\equiv& \bar{g}_{\mu\nu}dx^\mu dx^\nu \nonumber \\
&=&-\frac{\Delta}{\rho^2}\Big(dt -a \sin^2\theta d\phi\Big)^2 +\frac{\rho^2}{\Delta} dr^2+
\rho^2d\theta^2 +\frac{\sin^2\theta}{\rho^2}\Big[(r^2+a^2)d\phi -adt\Big]^2 \label{KBH}
\end{eqnarray}
with
\begin{eqnarray}
\Delta=r^2-2Mr+a^2,~ \rho^2=r^2+a^2 \cos^2\theta,~{\rm and}~a=\frac{J}{M}.
 \label{mps}
\end{eqnarray}
Here, $M$ and  $J$ represent the mass and angular momentum of the Kerr black hole.
Two  of  outer and inner horizons are found  by demanding $\Delta=(r-r_+)(r-r_-)=0$ as
\begin{equation}
r_{\pm}=M\pm \sqrt{M^2-a^2}.
\end{equation}
The line element (\ref{KBH}) is stationary but non-static because  $dt\to -dt$ changes the
signature of the metric and  it is also axially symmetric (invariance under $d\theta \to -d\theta$).

A  massive scalar perturbation $\Phi$  on the background of Kerr black holes is described  by
\begin{equation}
(\bar{\nabla}^2-\mu^2)\Phi=0.\label{phi-eq1}
\end{equation}
Considering  the axis-symmetric
background (\ref{KBH}), it is convenient to decompose the scalar perturbation
as
\begin{equation}
\Phi(t,r,\theta,\phi)=e^{-i\omega t+im\phi} S_{lm}(\theta)
 R_{lm}(r), \label{sep}
\end{equation}
where $S_{lm}(\theta)$ is the spheroidal harmonics  with $-m\le l
\le m$  and $R_{l m}(r)$ satisfies a radial part of the wave
equation.  Substituting (\ref{sep}) into (\ref{phi-eq1}), we have
an  angular  equation  and Teukolsky  equation for $R_{lm}(r)$~\cite{Hod:2012px}
\begin{eqnarray}
&& \frac{1}{\sin \theta}\partial_{\theta}\Big(
\sin \theta
\partial_{\theta} S_{lm}(\theta) \Big )+ \left [K_{lm}+a^2(\mu^2-\omega^2)\sin^2\theta-\frac{m^2}{\sin ^2{\theta}} \right ]S_{lm}(\theta) =0,
\label{wave-ang1}
\end{eqnarray}
\begin{eqnarray}
\Delta \partial_r \Big( \Delta \partial_r R_{l m}(r) \Big)+U(r)R_{l m}(r)=0,
\label{wave-rad}
\end{eqnarray}
where
\begin{eqnarray}
K_{lm}&=&l(l+1)-a^2(\mu^2-\omega^2)+\sum_{k=1}^{\infty}c_ka^{2k}(\mu^2-\omega^2), \label{K-coe} \\
U(r)&=&[\omega (r^2+a^2)-am]^2+\Delta[2am\omega-\mu^2(r^2+a^2) -K_{lm}].\label{U-pot}
\end{eqnarray}
Hereafter, we choose the separation constant $K_{lm}=l(l+1)+a^2(\mu^2-\omega^2)$ to be consistent with the geodesic potential $V_{\rm gK}(r)$ for a neutral particle (see  section 4), even though its lower bound is given by  $K_{lm}>l(l+1)-a^2(\mu^2-\omega^2)$~\cite{Berti:2005gp}. The coefficient $c_k$ appears up to $c_4$ in~\cite{Seidel:1988ue}.
It is worth noting that Eq. (\ref{wave-rad}) is usually  employed   to obtain  exact solutions.

Now, we need the tortoise coordinate $r_*$  to derive the Schr\"odinger-type equation as
\begin{equation}
r_*=\int\frac{(r^2+a^2)dr}{\Delta}=r+\frac{r_+^2+a^2}{r_+-r_-}\ln\Big(\frac{r}{r_+}-1\Big)-\frac{r_-^2+a^2}{r_+-r_-}\ln\Big(\frac{r}{r_-}-1\Big).
\end{equation}
Then,  the Teukolsky equation (\ref{wave-rad}) leads to
the Schr\"odinger-type equation when setting $\Psi_{l m}=\sqrt{r^2+a^2} R_{l m}$
\begin{equation}
\frac{d^2\Psi_{l m}(r_*)}{dr_*^2}+[\omega^2-V_{\rm K}(r)]\Psi_{l m}(r_*)=0, \label{sch-eq}
\end{equation}
where the scalar potential $V_{\rm K}(r)$ is found to be~\cite{Zouros:1979iw,Arvanitaki:2010sy}
\begin{eqnarray}
V_{\rm K}(r)&=&\omega^2\nonumber \\
&-&\frac{3\Delta^2r^2}{(a^2+r^2)^4}+\frac{\Delta[\Delta+2r(r-M)]}{(a^2+r^2)^3} \label{c-pot1} \\
&+&\frac{\Delta \mu^2}{a^2+r^2}-\Big(\omega-\frac{am}{a^2+r^2}\Big)^2
-\frac{\Delta}{(a^2+r^2)^2}\Big(2am\omega -l(l+1)-a^2(\mu^2-\omega^2)\Big). \nonumber
\end{eqnarray}
Here,  the second line of Eq.(\ref{c-pot1}) represents the effect of introducing the tortoise coordinate $r_*$, while the last line  comes form $-U(r)/(r^2+a^2)^2$ in Eq.(\ref{wave-rad}).
The latter part determines the near-horizon  and asymptotic limits: $V_{\rm K}(r\to r_+) = \omega^2-(\omega-\omega_c)^2$ with the critical frequency $\omega_c=m a/(r_+^2+a^2)$ and
$V_{\rm K}(r\to \infty)=\mu^2$.
Taking the asymptotic limit of Eq. (\ref{sch-eq}) and its near-horizon limit, one has the plane-wave solutions
\begin{eqnarray}
\Psi^\infty(r)&\sim&  e^{-i\sqrt{\omega^2-\mu^2} r_*}(\leftarrow)+{\cal R}e^{i\sqrt{\omega^2-\mu^2} r_*}(\rightarrow),\quad r_*\to +\infty(r\to \infty), \label{asymp1}\\
\Psi^{-\infty}(r)&\sim& {\cal T} e^{-i(\omega-\omega_c) r_*}(\leftarrow),\quad r_*\to -\infty(r\to r_+), \label{asymp2}
\end{eqnarray}
where  ${\cal T}({\cal R})$ are the transmission (reflection) amplitudes.

Imposing the flux conservation, we obtain the relation between reflection and transmission coefficients as
\begin{equation}
|{\cal R}|^2=1-\frac{\omega-\omega_c}{\sqrt{\omega^2-\mu^2}}|{\cal T}|^2,
\end{equation}
which  means that only waves with $\omega>\mu$ propagate to infinity and the superradiant scattering  may occur ($\rightarrow,~|{\cal R}|^2>|{\cal I}|^2$) whenever $\omega<\omega_c$ (superradiance condition) is satisfied  because  outgoing waves at the outer horizon reinforce the outgoing waves at infinity.
On the other hand, one may choose the scalar modes to have an exponentially decay as it tends to zero  at infinity
 \begin{equation}
\Psi^{\rm b,\infty} \sim  e^{-\sqrt{\mu^2-\omega^2} r} \rightarrow 0 \label{asymp-b}
 \end{equation}
together  with the bound state condition of  $\omega<\mu$.

At this stage, we wish to mention   three cases  for a  massive  scalar propagating around the Kerr black holes based on the potential analysis:\\
Case (i) superradiant stability: $\omega<\omega_c$ and  $\omega<\mu$  without a positive trapping well. \\
Case (ii) stationary resonances (marginally stable)~\cite{Hod:2012px}: $\omega=\omega_c$ and $\omega<\mu$.\\
Case (iii) superradiant instability: $\omega<\omega_c$ and  $\omega<\mu$  with  a positive trapping well.\\

When solving Eq.(\ref{wave-rad}) directly, the frequency $\omega $ is permitted to be complex (small complex modification)
 as~\cite{Dolan:2007mj}
\begin{equation}
\omega=\omega_{\rm R}+i \omega_{\rm I}.
\end{equation}
In this case, the sign of $\omega_{\rm I}$ determines the solution which  is decaying ($\omega_{\rm I}<0$) or growing ($\omega_{\rm I}>0$) in time. Here, one describes again  the above cases: \\
Case (i): $\omega_{\rm I}<0$ and  $\omega_{\rm R}>\omega_c$. The  solution is stable (decaying in time). \\
Case (ii): $\omega_{\rm I}=0$  and $\omega_{\rm R}=\omega_c$. \\
Case (iii): $\omega_{\rm I}>0$ and $\omega_{\rm R}<\omega_c$.  The  solution is unstable (growing in time).\\
Case (i) is the only one that is present for Schwarzschild black holes~\cite{Damour:1976kh,Zouros:1979iw,Detweiler:1980uk,Dolan:2007mj}.
Case (ii) corresponds to bound states of marginally stable modes~\cite{Hod:2012px}. It is easy to understand the existence of such stationary solutions because their frequency saturates the superradiant condition.

Case (iii) corresponds to superradiant instability  of Kerr black holes under a massive scalar propagation. The mode is bounded,  the wave function is peaked far outside the outer horizon, and $\omega$ is nearly real with $|\omega_{\rm I}|\ll \omega_{\rm R}$.
For large mass $M\mu \gg1$, $m=l$, $a/M\simeq 1$, and $\omega_{\rm R} \sim 0.98\mu<\omega_c$ whose potential is given by Fig.4,  Zouros and Eardley  have found the maximum growth rate ($M\omega_{\rm I}\sim 10^{-7} e^{-1.84 M \mu}$) by using the WKB method~\cite{Zouros:1979iw}. On the other hand, for small mass $M\mu\ll1$, $m=l=1$, $a/M\simeq 1$, and $\omega_{\rm R}\sim \mu$, Detweiler has found the maximum growth rate $M\omega_{\rm I}\sim a(M\mu)^9/(24M^2)$. To obtain this rate, one  approximated $R_{lm}(r)$ by known analytic functions in two asymptotic regions~\cite{Detweiler:1980uk}: the near horizon wave function is  the hypergeometric function ${}_2F_1(a,b,c;x)$ and the far-region wave function is given by confluent hypergeometric function $U(a,b;cr)$ with $k=l+1/2$ in Eq.(\ref{wavef-1}).
For $M\mu(=0.42)\le 0.5$, $m=l=1$, and $a=0.99$, Dolan has obtained a maximum growth rate of $M\omega_{\rm I} \sim 1.5 \times 10^{-7}$ by making use of a continued-fraction method adopted for computing quasinormal modes~\cite{Dolan:2007mj}.

\section{Potential analysis}
\begin{figure*}[t!]
   \centering
  \includegraphics{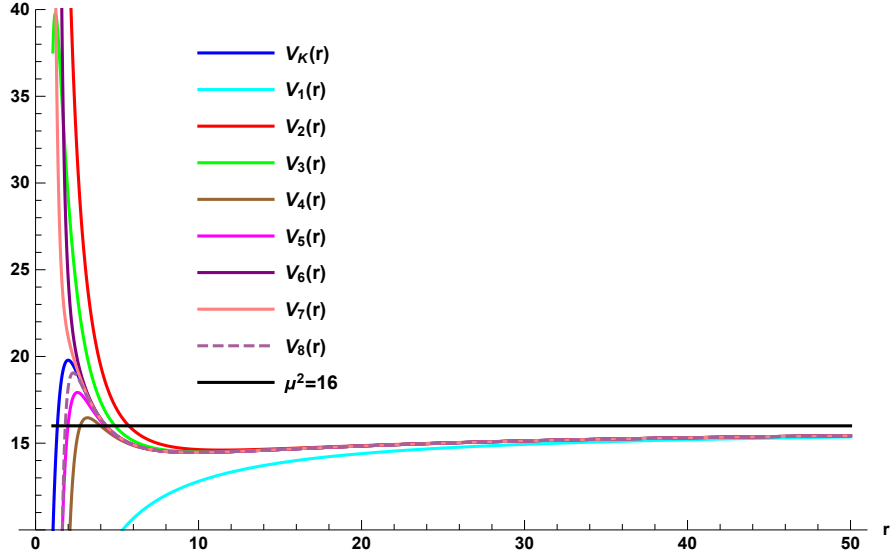}
\caption{ Full potential $V_{\rm K}(r)$ and  potential $V_i(r),\{i=1,\cdots,8\}$ as functions of $r\in[r_+ =1.063,50]$ with $M=1,\omega=3.85,a=0.998,m=13,l=13,\mu=4$. All potentials except $V_1(r)$ show a trapping well (local minimum in the region of $9.8<r<11.4$).  }
\end{figure*}
First of all, we derive the potential $V_i(r)$ when expanding $V_{\rm K}(r)$ in Eq.(\ref{c-pot1}) for large $r$ as
\begin{eqnarray}
V_{i}(r)&=&\mu^2-\frac{2M\mu^2}{r} \nonumber \\
&+&\frac{\ell(\ell+1)+a^2(\mu^2-\omega^2)}{r^2} \nonumber \\
&+& \frac{2M(1- l(l+1)+2ma\omega+a^2\omega^2)}{r^3}\nonumber \\
&-&\frac{4M^2+a^2(m^2-1+l(l+1))+a^4(\mu^2-\omega^2)}{r^4}\label{p-exp} \\
&+& \frac{4Ma^2(-3+l(l+1))-2a^3 M(2a\omega^2+4m\omega-a\mu^2)}{r^5}\nonumber \\
&+& \frac{24 M^2a^2+a^4(2m^2-2+l(l+1))+a^6(\mu^2-\omega^2)}{r^6}\nonumber \\
&+& \frac{6Ma^4(5-l(l+1))+2Ma^5(-2a \mu^2+6m \omega +3a\omega^2)}{r^7} \nonumber \\
&-& \frac{72M^2a^4+a^6(3m^2-3+l(l+1))+a^8(\mu^2-\omega^2)}{r^8} \nonumber \\
&+& {\cal O}\Big(\frac{1}{r^9}\Big).\nonumber
\end{eqnarray}
We label $V_i(r)$ for up to $(1/r)^i$-order with $i=1,2,3,4,5,6,7,8$. As is shown in Fig. 1, $\{V_2(r),V_3(r),V_6(r),V_7(r)\}$ are increasing functions, while $\{V_{\rm K}(r),V_1(r),V_4(r),V_5(r),V_8(r)\}$ are decreasing functions in the near horizon. All potentials converge into $\mu^2=16$ in the asymptotic region and they describe superradiant instability (see Fig. 4).
Also, all potentials except $V_{1}(r)$ indicate a local minimum in the region of $9.8<r<11.4$, which implies that $V_2(r)$ is the lowest order potential to represent a trapping well.
$V_1(r)$ and $V_2(r)$  are  relevant to analyzing superradiant instability~\cite{Myung:2022kex,Myung:2022biw}.
An  asymptotic  form of the Newton potential is given by
\begin{equation}
V_1(r)=\mu^2-\frac{2M\mu^2}{r} \label{asym-p}
\end{equation}
which appears  in the asymptotic region. We stress that $V_1(r)$ is compared to $V_{\rm asy}(r)=\mu^2+(2M\mu^2-4M\omega^2)/r$ in~\cite{Hod:2012zza} where the attractive  Newtonian term is absent.
This potential may be used to find the condition for a trapping well as
\begin{equation}
V'_1(r)>0 \quad  \to M\mu^2>0. \label{cond-t}
\end{equation}
However, the condition  ($V'_1(r)<0$) for no trapping well is not  allowed  because $V'_1(r)=2M\mu^2/r^2$ is always positive.
It is worth noting that  Eq.(\ref{cond-t}) is not a sufficient condition to get  a trapping well.
We have to find the other condition. For this purpose,
we introduce  the far-region potential appeared in the large $r$ region
\begin{equation}
V_2(r)=\mu^2-\frac{2M\mu^2}{r}+\frac{\ell(\ell+1)+a^2(\mu^2-\omega^2)}{r^2}, \label{fr-p}
\end{equation}
where the last term plays a crucial role of making a trapping well. It is interesting to  note that $V_2(r)$ matches $V_{\rm gK}(r)$ (\ref{g-pot}) with $L^2=l(l+1)$ exactly up to $(1/r^2)$-order. If $a^2$-term is absent, $V_2(r)$ is identical with the asymptotic  potential obtained  for $M\omega \ll 1$, $M\mu\ll 1$, and $r\gg M$ in Ref.~\cite{Detweiler:1980uk}.
We note that an  effect of $a^2$-term is  limited because of $a^2\le 1$ with $ M=1$.  This can be  seen easily from  a centrifugal potential ($l(l+1)/r^2$-term)  which has a greatly repulsive effect on making a trapping well for large $l$. Even though  $a^2$-term is less effective than  centrifugal potential, one should include  it. This is  because $V_2(r)$ reduces to that for the Schwarzschild black hole when neglecting it.     It is important to note that there is no way to make a trapping well (a positive local minimum)  if one keeps the Newton potential  $V_1(r)$ only.
Here, we  have an extremal point ($r_{\rm e}$)
\begin{equation}
V'_2(r_{\rm e})=0 \quad \to r_{\rm e}=\frac{l(l+1) +a^2(\mu^2-\omega^2)}{M\mu^2}>r_+,
\end{equation}
which becomes a local minimum, located far from the outer horizon when the bound is satisfied as
\begin{equation}
l(l+1)>-a^2(\mu^2-\omega^2).  \label{l-bound}
\end{equation}

\begin{figure*}[t!]
   \centering
  \includegraphics{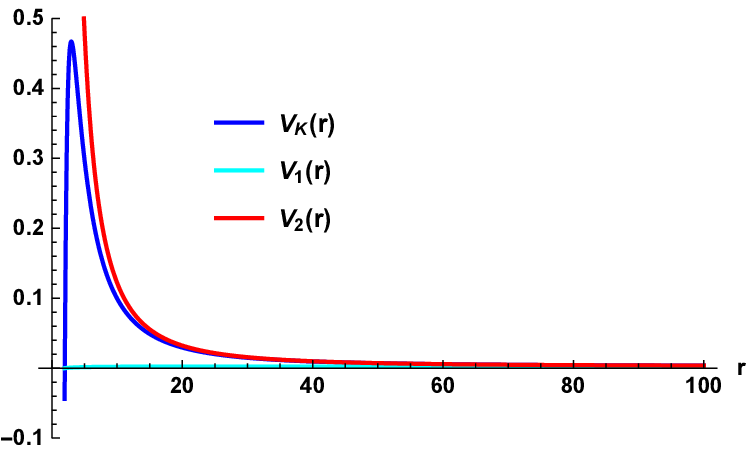}
   \hfill%
  \includegraphics{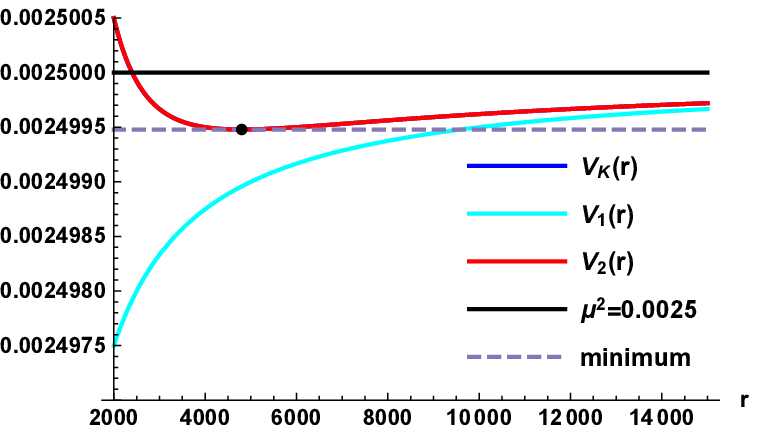}
\caption{(Left) Superradiant stable  potential $V_{\rm K}(r)$   as function of $r\in[r_+ =1.95,100]$ with $M=1,\omega=0.02,a=0.3,m=3,\ell=3,\mu=0.05$.
We have a negative potential $V_{\rm K}(r_+)=-0.044$.
  We check the conditions of $\omega<\mu$ and $\omega<\omega_c(=0.08)$ to have a superradiant stability, but $V'_1(r)>0$ predicts a well. (Right) Asymptotic forms of  $V_{\rm K}(r)\simeq V_2(r)$ indicate a tiny well ($\bullet$) located at $r=r_e=4800$. $V_1(r)$ approaches them for $r>r_e$. }
\end{figure*}

Hereafter, we choose $M=1$ so that $r$ and $a$ are measured in units of $M$, while $\omega$ and $\mu$ are measured in units of $M^{-1}$.
It is curious to note that (Left) Fig. 2 corresponds to a  superradiantly stable potential  because we could not find a trapping well  for $\omega<\omega_c=0.08$ and $\omega<\mu$.
In this Case (i), however, we observe   $V'_1(r)>0$ which may imply the superradiant instability. So,  $V'_1(r)>0$ contradicts to our expectation of no trapping well.
We wish to resolve it. We find from (Right) Fig. 2 that a tiny  well is located at a very large distance of $r=4800$ in $V_{\rm K}(r)\simeq V_2(r)$. It shows that $V'_1(r)>0$ implies either a trapping well or a tiny well. Hence, one has to find the other condition for a trapping well in the next section.
\begin{figure*}[t!]
   \centering
  \includegraphics{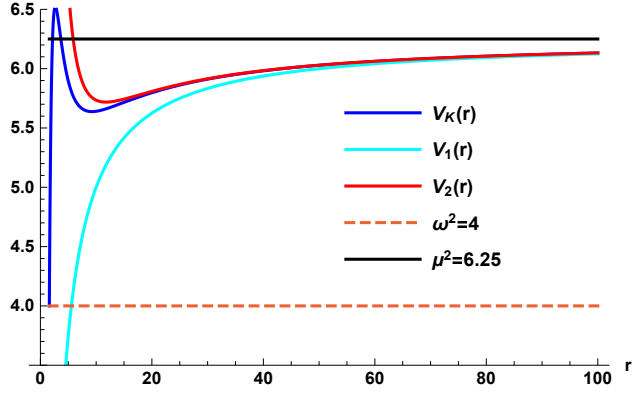}
\caption{Stationary  resonances   potential $V_{\rm K}(r)$   as function of $r\in[r_+ =1.6,100]$ with $M=1,\omega=2,a=0.8,m=8,l=8,\mu=2.5$.
We have $V_{\rm K}(r_+)=4=\omega^2$ and $V_{\rm K}(r_e=9.2)=5.6(>\omega^2)$.
We note   $\omega=\omega_c=2$ and $\omega<\mu$ to meet the condition for stationary scalar clouds.   }
\end{figure*}

To visualize stationary resonances [Case (ii)], we observe the corresponding potential $V_{\rm K}(r)$ with $a=0.8$ and $\mu=2.5$ in Fig. 3.
This  is similar apparently  to the superradiant instability because it has a trapping well except $\omega=\omega_c$. But, imposing $\omega=\omega_c$ will be affected  in the near-horizon and far-region regions.

\begin{figure*}[t!]
   \centering
  \includegraphics{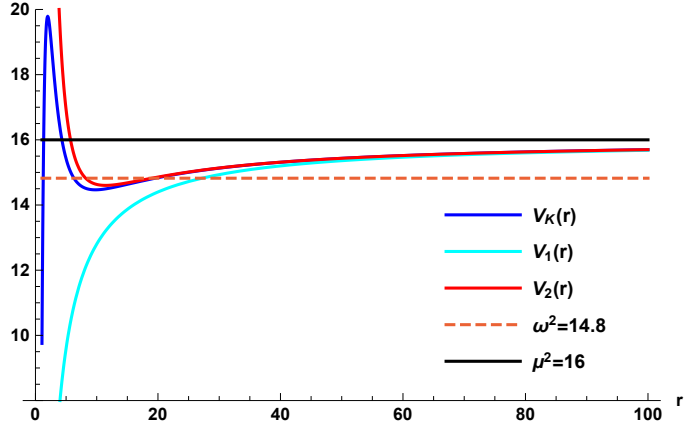}
\caption{Potential $V_{\rm K}(r)$   as function of $r\in[r_+ =1.063,100]$ with $M=1,\omega=3.85,a=0.998,m=13,l=13,\mu=4$. A local minimum of $V_{\rm K}(r_e)(=14.5)<\omega^2$ is located at $r=r_e=9.8$.
This potential is a standard type of ergo-barrier-well-mirror with $V_{\rm K}(r_+)=9.75$. We note  $\omega<\mu$ and $\omega<\omega_c(=6.1)$ with a trapping well for superradiant instability. }
\end{figure*}
\begin{figure*}[t!]
   \centering
  \includegraphics{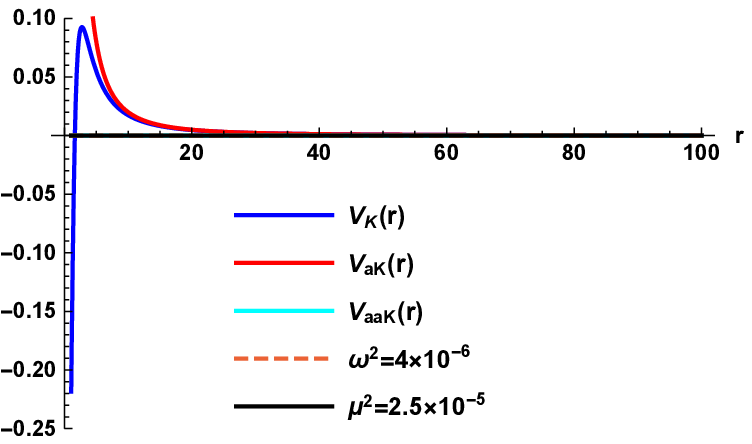}
   \hfill%
  \includegraphics{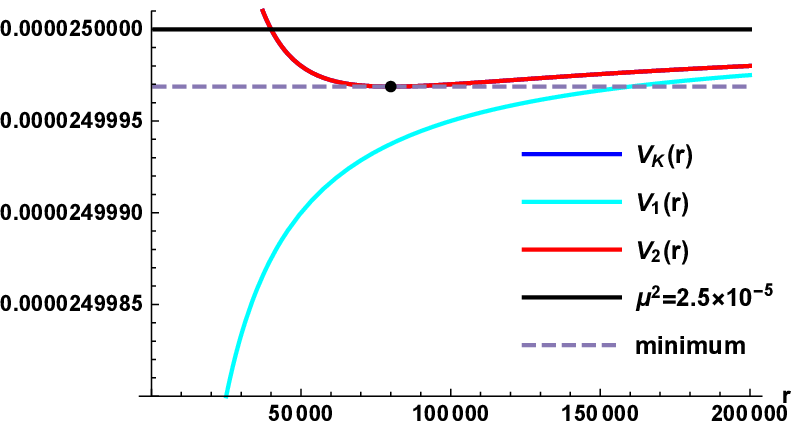}
\caption{(Left) Potential $V_{\rm K}(r)$   as function of $r\in[r_+ =1.063,100]$ with $M=1,\omega=0.002,a=0.998,m=1,l=1,\mu=0.005$.
One has a negative potential $V_{\rm K}( r_+)=-0.22$ and a peak of $V_{\rm K}(r)$ is located at $r=2.76$.
This represents  an  ultralight boson with $\omega<\omega_c(=0.47)$.
   (Right) Asymptotic  forms of  $V_{\rm K}(r)\simeq V_2(r)$ with $V'_1(r)>0$ indicate a tiny well  located at asymptotic region ($\bullet$ at $r=r_e=8.0\times 10^4$). }
\end{figure*}
For Case (iii), let us display the potential Eq.(\ref{c-pot1}) with $a=0.998$ and $\mu=4$ in  Fig. 4 (Fig. 1) which implies the superradiant instability~\cite{Myung:2022kex}.
This potential has a typical shape of ergo-barrier-well-mirror and the trapping well plays an essential role in achieving the superradiant instability~\cite{Zouros:1979iw,Arvanitaki:2010sy,Konoplya:2011qq}.
In this case, one has $\omega<\mu$ for having  asymptotic bound states  and $\omega<\omega_c$ for superradiant states (outgoing waves at the outer horizon).

On the other hand, (Left) Fig. 5 indicates a potential for an ultralight boson with $\mu=0.005 \ll 1/M$ and $V_{\rm K}(r)\simeq V_2(r)$ have  a tiny well located at asymptotic region ($r=r_e=8.0\times 10^4$) [(Right) Fig. 5]. Interestingly, we observe a negative  potential at the outer horizon [$V_{\rm K}(r_+)=-0.22$].

Lastly, we propose an increasing  potential appeared in  Fig. 6 which represents a boundary between  trapping well and  tiny well. In this case, we have $V_{\rm K}(r)\simeq V_1(r) \simeq V_2(r)$ without  extremal points outside the outer horizon because  a tiny well is located at $r=r_e=1.13 \times 10^{15}$.  Also, it violates the superradiant condition of $\omega<\omega_c$.

\begin{figure*}[t!]
   \centering
  \includegraphics{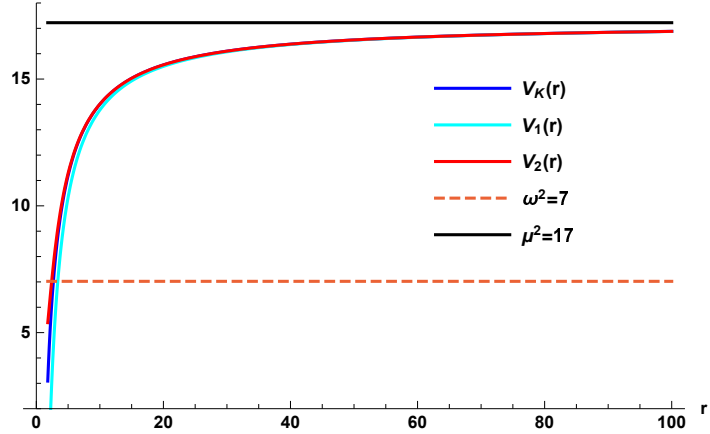}
\caption{Potential $V_{\rm K}(r)\simeq V_1(r) \simeq V_2(r)$   as functions of $r\in[r_+ =1.8,100]$ with $M=1,\omega=2.65,a=0.6,m=4,l=4,\mu=4.15$.
They all are increasing potentials without any extremal point $r=r_e$. We note  $\omega<\mu$ and $\omega>\omega_c(=0.17)$. }
\end{figure*}
According to the potential analysis for the Kerr black holes, we have either a trapping well or a tiny well because of $V'_1(r)>0$ in the Kerr black hole background.

\section{Geodesic potential}

In this section, we  derive a geodesic potential to compare with $V_{\rm K}(r$) and $V_3(r)$.
We consider the Lagrangian for a neutral particle moving around  the Kerr geometry~\cite{Ivanov:2005cf,Liu:2017fjx}
\begin{equation}
{\cal L}=\frac{1}{2}\bar{g}_{\rm \mu\nu}\frac{dx^\mu}{d\lambda}\frac{dx^\nu}{d\lambda}
\end{equation}
with $\lambda=\tau/\tilde{m}$ the proper time per unit mass.
The radial equatorial ($\theta=\pi/2$) motion for a neutral particle  takes the form
\begin{equation}
\Big(\frac{dr}{d\lambda}\Big)^2=E^2-V_{\rm gK}(r),
\end{equation}
where $E=\partial {\cal L}/\partial \dot{t}$ is the conserved energy of the particle and $V_{\rm gK}(r)$ is the the geodesic potential defined as
\begin{eqnarray}
V_{\rm gK}(r)= \tilde{m}^2-\frac{2M\tilde{m}^2}{r} +\frac{L^2+a^2(\tilde{m}^2-E^2)}{r^2}
+\frac{2M(-L^2+2aLE-a^2E^2)}{r^3}. \label{g-pot}
\end{eqnarray}
Here,  $L=\partial {\cal L}/\partial \dot{\phi}$ is the conserved projection of the   particle's angular momentum on the axis of the black hole.
Considering the correspondence of $\tilde{m}\leftrightarrow \mu$,  $E\leftrightarrow\omega$, $L\leftrightarrow m$,  and $L^2 \leftrightarrow l(l+1)$, the geodesic potential $V_{\rm gK}(r)$ matches the scalar potential $V_{\rm 3}(r)$  when neglecting the first term in the last line of $V_3(r)$ which comes from  introducing the tortoise coordinate $r_*$. One  point mismatched  is the sign of the last term in $(1/r^3)$-term.
We note that $V_2(r)$ matches $V_{\rm gK}(r)$ exactly up to $(1/r^2)$-order.
This explains why we choose $K_{lm}=l(l+1)+a^2(\mu^2-\omega^2)$ in this work.
\begin{figure*}[t!]
   \centering
  \includegraphics{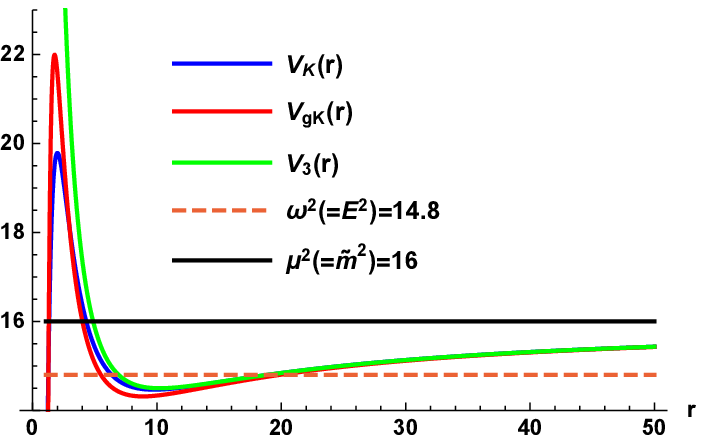}
   \hfill%
  \includegraphics{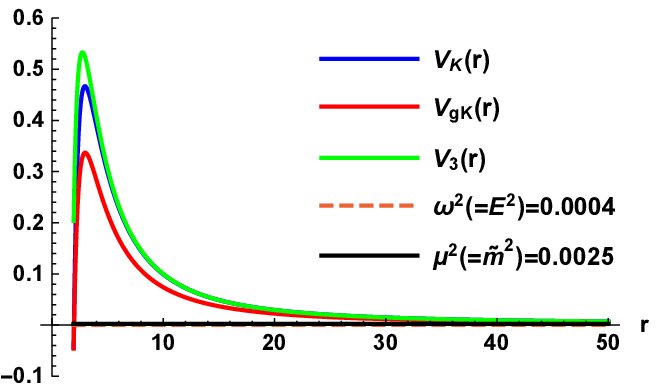}
\caption{(Left) Three potentials $V_{\rm K}(r)$, $V_{\rm gK}(r)$, and $V_3(r)$    as functions of $r\in[r_+ =1.063,100]$ with $M=1,\omega(=E)=3.85,a=0.998,m=13,l(=L)=13,\mu(=\tilde{m})=4$. $V_{\rm K}(r)$ appeared in Fig. 4.
Local minima of  $V_{\rm K}(r)$, $V_{\rm gK}(r)$, and $V_3(r)$ are located at $r=9.8,8.9,10.2$, respectively.
   (Right) Three potentials $V_{\rm K}(r)$, $V_{\rm gK}(r)$, and $V_3(r)$    as functions of $r\in[r_+ =1.95,100]$ with $M=1,\omega(=E)=0.02,a=0.3,m=3,l(=L)=3,\mu(=\tilde{m})=0.005,L=3$. $V_{\rm K}(r)$ appeared in Fig. 2. We have $V_{\rm K}(r_+)=-0.044$, $V_{\rm gK}(r_+)=-0.046$, and  $V_3(r_+)=0.2$ at the outer horizon. }
\end{figure*}
We display  $V_{\rm gK}(r)$, $V_{\rm K}(r)$,  and $V_3(r)$ in Fig. 7.
 The maximum (minimum) of the geodesic potential $V_{\rm gK}(r)$ in (Left) Fig. 7 is associated with unstable (stable) circular geodesic  orbit. In turn, the unstable (stable ) circular orbit is related to the low overtones of the quasinormal mode (quasibound state) spectrum via an association  between $V_3(r)$ and $V_{\rm gK}(r)$ in the eikonal regime $(l+1/2\gg1)$~\cite{Percival:2020skc}.

\section{Far-region and asymptotic wave functions}

It is important to find  the scalar wave forms in the  far-region  to distinguish between quasibound state (trapping well)  and  bound state (tiny well).
This is so  because the condition of $V'_1(r)>0$ in Eq.(\ref{cond-t}) is always satisfied and it is not a sufficient condition to get a trapping well.
Now, we  use $V_2(r)$ to obtain far-region wave functions.
In the far-region where one  may choose $r_*\simeq r$, an equation from  (\ref{sch-eq}) together with (\ref{fr-p}) takes the form
\begin{equation}
\Big[\frac{d^2}{dr^2}+\omega^2-V_2(r)\Big]\Psi_{lm}(r)=0 \label{far-eq}
\end{equation}
whose solution is given exactly by the confluent  hypergeometric function $U(a,b;cr)$ as
\begin{eqnarray}
\Psi_{\ell m}(r)&=&c_1 e^{-\sqrt{\mu^2-\omega^2}r} \Big(2\sqrt{\mu^2-\omega^2} r\Big)^{k+\frac{1}{2}} \nonumber \\
 &\times & U\Big(k+\frac{1}{2}-\frac{M\mu^2}{\sqrt{\mu^2-\omega^2}},2k+1;2\sqrt{\mu^2-\omega^2} r\Big) \label{wavef-1}
\end{eqnarray}
with
\begin{equation}
k=\frac{1}{2}\sqrt{1+4[\ell(\ell+1)+a^2(\mu^2-\omega^2)]}.
\end{equation}
Here, we note that $k$ is real if Eq.(\ref{l-bound}) holds.
Also,  we observe a bound state of  $e^{-\sqrt{\mu^2-\omega^2}r}$ with $\omega<\mu$ appeared  in (\ref{asymp-b}).
Furthermore, we find  some information from the large $r$-expansion of $U(a,b;cr)$ as
\begin{equation}
U(a,b;cr\to \infty)\rightarrow\quad  \frac{1}{(cr)^{a}}\Big[1-\frac{a(1+a-b)}{cr} +{\cal O}\Big(\frac{1}{cr}\Big)^2\Big], \label{large-U}
\end{equation}
which implies roughly that one finds  a decreasing function $U(a,b;cr)$ for a positive $a$, while one has an increasing function $U(a,b;cr)$ for a negative $a$.
Plugging  Eq.(\ref{large-U}) into Eq.(\ref{wavef-1}) leads to the asymptotic wave function
as
\begin{equation}
\Psi^{\rm A}(r)\simeq e^{-\sqrt{\mu^2-\omega^2}r} \Big(2\sqrt{\mu^2-\omega^2} r\Big)^{\frac{M\mu^2}{\sqrt{\mu^2-\omega^2}}}, \label{wavef-2}
\end{equation}
which is exactly the same solution obtained  if one includes $V_1(r)$ in Eq.(\ref{far-eq}), instead of $V_2(r)$.
\begin{figure*}[t!]
   \centering
  \includegraphics{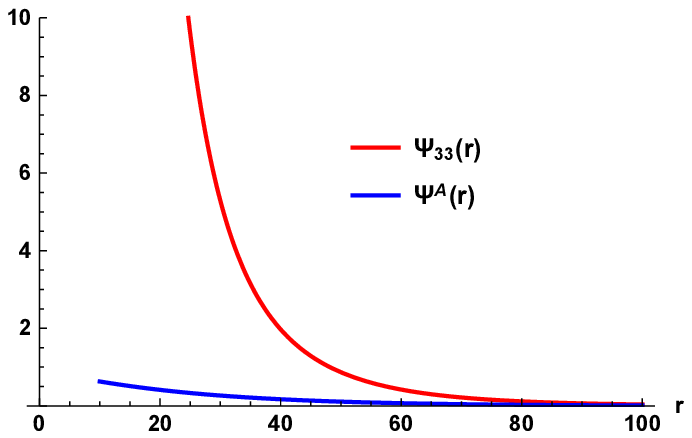}
  \hfill%
  \includegraphics{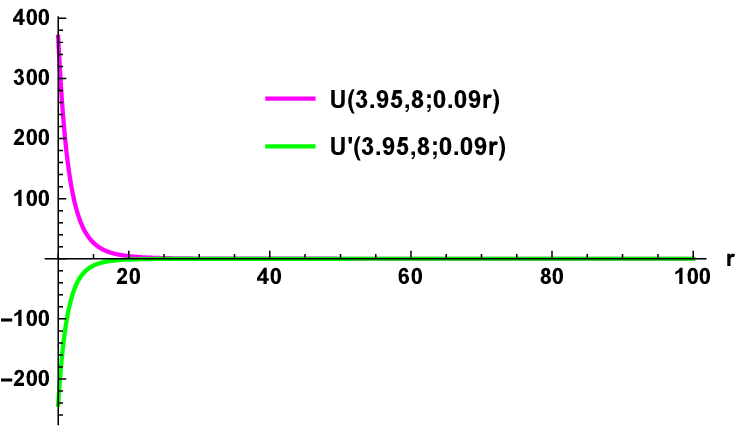}
\caption{ (Left) Bound state function $\Psi_{33}(r)$ and its asymptotic wave function $\Psi^{\rm A}(r)$ corresponding to Fig. 2
 (Right) Confluent hypergeometric function $U(3.95,8;0.09r)$ as $r\in[10,100]$ is decreasing,  while its first derivative $U'(3.95,8;0.09r)$ is negative. }
\end{figure*}

Considering the potential  without trapping well [(Right) Fig. 2], $\Psi_{33}(r)$ in Fig. 8 shows a bound state. The asymptotic wave function $\Psi^{\rm A}(r)$ is a decreasing function.  Also, we have a rapidly decreasing function  $U(3.95,8;0.09r)$ and its first derivative $U'(3.95,8;0.09r)$ is negative. This case represents no trapping well clearly. Even though the corresponding  potential includes a tiny well located at $r=4800$ [see (Right) Fig. 2], one could not find any quasibound state [see (Left) Fig. 14].
\begin{figure*}[t!]
   \centering
  \includegraphics{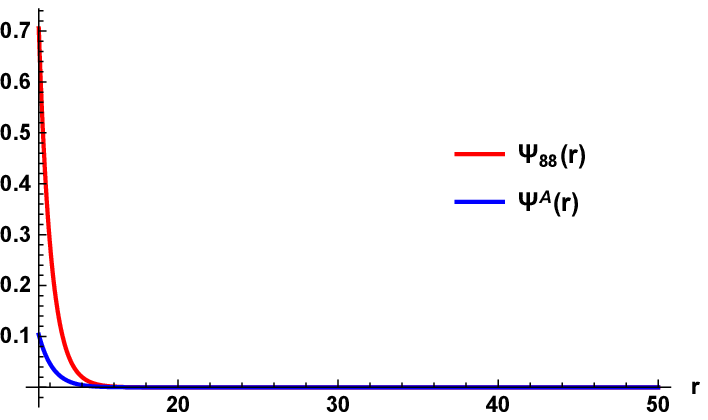}
  \hfill%
  \includegraphics{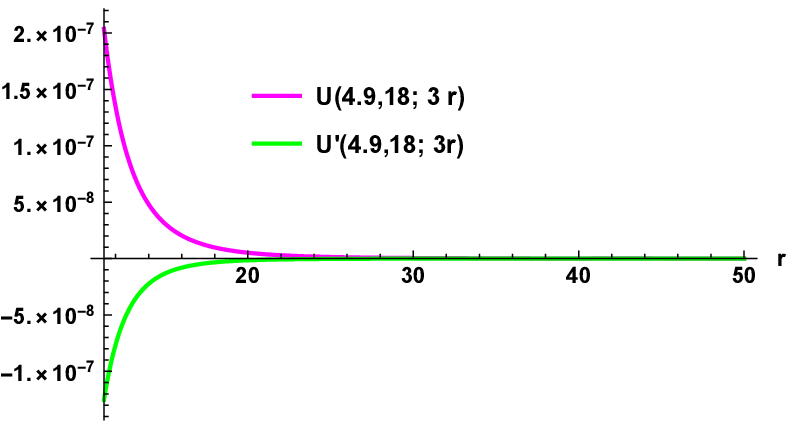}
\caption{(Left) Bound state function $\Psi_{88}(r)$ and its asymptotic wave function $\Psi^{\rm A}(r)$ corresponding to stationary resonances in Fig. 3.
 (Right) Confluent hypergeometric function $U(4.9,18;3r)$ as $r\in[10,100]$ is decreasing,  while its first derivative $U'(4.9,18;3r)$ is negative. }
\end{figure*}

Observing a potential representing  stationary resonances [Fig. 3], the wave function $\Psi_{88}(r)$ in (Left) Fig. 9 shows a bound state. The asymptotic wave function $\Psi^{\rm A}(r)$ is an exponentially decreasing function.  Also, we have a decreasing function  $U(4.9,18;3r)$ and its first derivative $U'(4.9,18;3r)$ is negative. This case represents no trapping well clearly, even though its potential has a trapping well.

\begin{figure*}[t!]
   \centering
  \includegraphics{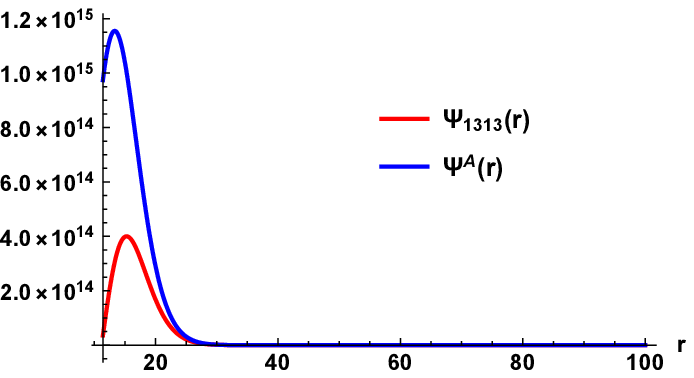}
  \hfill%
  \includegraphics{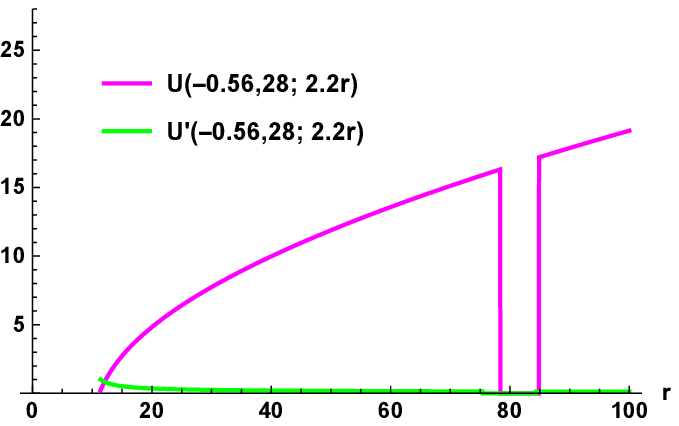}
\caption{(Left) Quasi-bound state of $\Psi_{1313}(r)$ as function of $r\in[11.4,100]$ with a trapping well [Fig. 4]. Here, we start with $r=11.4$ because a local minimum of $V_2(r)$ is located at $r=11.4$. (Right) Confluent hypergeometric function $U(-0.56,28;2.2r)$ represents  an increasing function and its derivative $U'(-0.56,28;2.2r)$ is positive.}
\end{figure*}
Let us observe the wave function  $\Psi_{1313}(r)$ for a trapping well [see Fig. 4].
As is shown in (Left) Fig. 10, $\Psi_{1313}(r)$ shows  a quasibound state (peak).
In this case, one has an increasing function $U(-0.56,28;2.2r)$ [(Right) Fig. 10] where a square well around $r=80$ is not important, while its first derivative $U'(-0.56,28;2.2r)$  is positive.
\begin{figure*}[t!]
   \centering
  \includegraphics{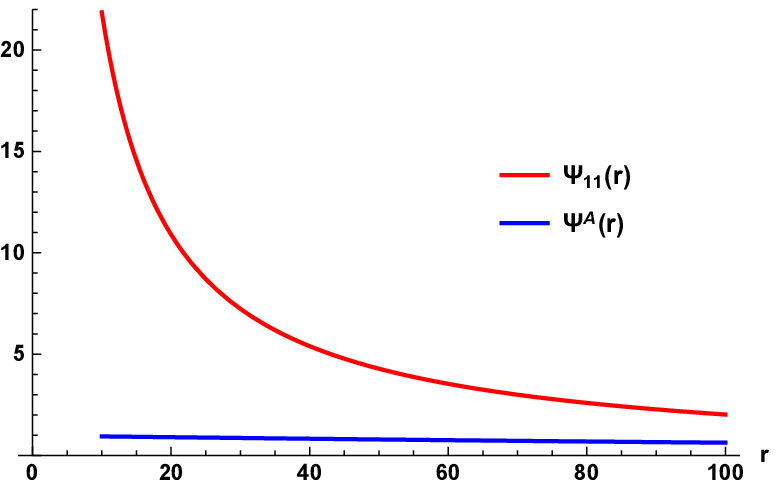}
\caption{Bound state function $\Psi_{11}(r)$  as $r\in[10,100]$ for an ultralight boson [(Left) Fig.5]. Its asymptotic wave function $\Psi^{\rm A}(r)$ represents a slowly decreasing function.}
\end{figure*}
\begin{figure*}[t!]
   \centering
  \includegraphics{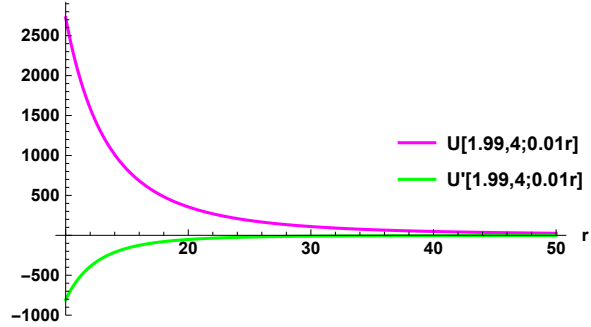}
\caption{Confluent hypergeometric function $U(1.99,4;0.01r)$ as $r\in[10,100]$ is decreasing,  while its first derivative $U'(1.99,4;0.01r)$ is negative. This picture  is designed for an ultralight boson. }
\end{figure*}

Now, we are in a position to introduce  an interesting  potential for an ultralight boson [(Left) Fig. 5].  The corresponding wave function $\Psi_{11}(r)$ [Fig. 11] is a decreasing function  and its asymptotic wave function $\Psi^{\rm A}(r)$ is a slowly decreasing function.
The confluent hypergeometric function $U(1.99,4;0.01r)$ [Fig. 12] indicates a  decreasing function and  the first derivative $U'(1.99,4;0.01r)$  is negative.
  Although this potential includes a tiny well located at $r=8.0 \times 10^4$ [see (Right) Fig. 5], one could not find a quasibound state [see (Right) Fig. 14]. This may contradict to our expectation that an ultralight boson could have superradiant instability.
\begin{figure*}[t!]
   \centering
   \includegraphics{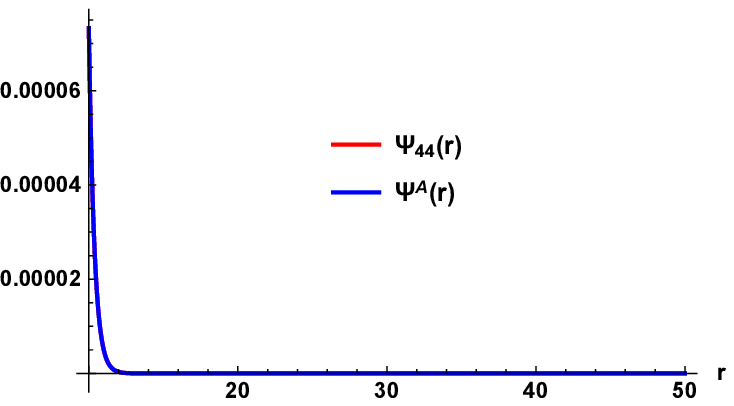}
  \hfill%
  \includegraphics{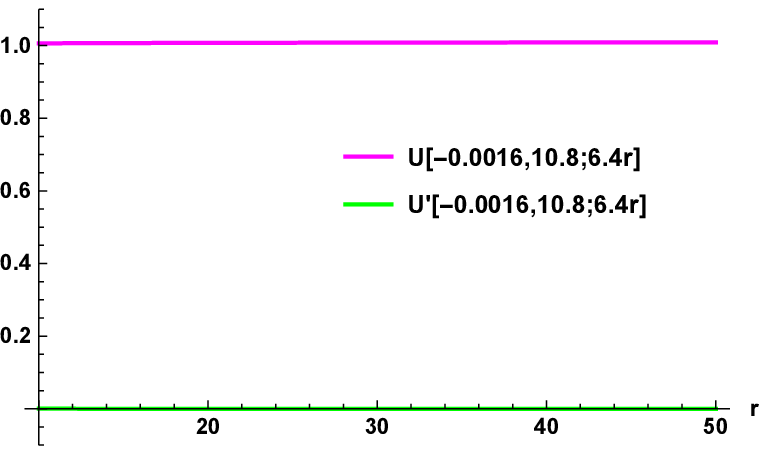}
\caption{ (Left) $\Psi_{44}(r)\simeq \Psi^{A}(r) $ represent a half of a peak (quasibound state).
(Right) Constant hypergeometric function $U(-0.0016,10.8;6.4r)$ and its derivative $U'(-0.0016,10.8;6.4r)\simeq 0$ for Fig. 6.}
\end{figure*}

Finally, we observe the increasing  potential [Fig. 6]. In this case,  one has a half of a peak (quasibound state), $\Psi_{44}(r)\simeq\Psi^{\rm A}(r)$ [see (Left) Fig. 13].
As is shown in (Right) Fig. 13, the confluent hypergeometric function $U(-0.0016,10.8;6.4r)$ is constant nearly and thus, its first derivative  $U'(-0.0016,10.8;6.4r)$ is zero nearly ($a\simeq 0$). This case may represent a boundary between trapping well ($a<0$) and  tiny well ($a>0$).

Therefore, the quasibound state  could be achieved when the first argument of $U(a,b;cr)$ is negative as
\begin{equation}
a<0 \to \quad \frac{M\mu^2}{\sqrt{\mu^2-\omega^2}}>k+\frac{1}{2} \label{trap-well}
\end{equation}
which is considered as  the condition to get a trapping well.
 On the other hand,
the bound state   could be found  when the first argument of $U(a,b;x)$ is positive as
\begin{equation}
a>0 \to \quad \frac{M\mu^2}{\sqrt{\mu^2-\omega^2}}<k+\frac{1}{2},\label{no-tw}
\end{equation}
which is regarded as the condition for  no trapping well (a tiny well).
\begin{figure*}[t!]
   \centering
   \includegraphics{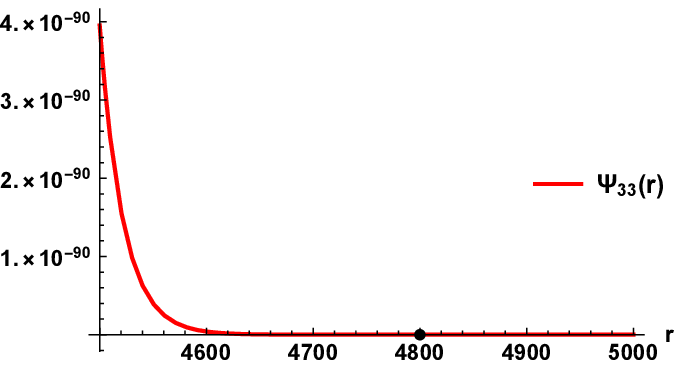}
  \hfill%
  \includegraphics{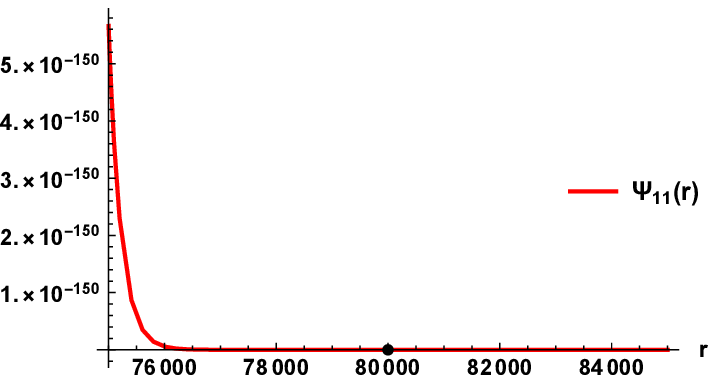}
\caption{ (Left) Asymptotic scalar function  $\Psi_{33}(r) $ as a function of $r\in[4500,5000]$  represents an asymptotic  bound state. This  includes a tiny well ($\bullet$) at $r=4800$, but
it has nothing special.
(Right) Asymptotic scalar function  $\Psi_{11}(r) $ as a function of $r\in[75000,85000]$  represents an asymptotic bound state. Although this  includes a tiny well ($\bullet$) at $r=8.0\times 10^4$,
this point is nothing special. }
\end{figure*}
\section{Discussions}
First of all, we have obtained the massive  scalar potential $V_{\rm K}(r)$  with separation constant $K_{lm}=l(l+1)+a^2(\mu^2-\omega^2)$ in the Kerr black hole background.
This separation constant was fixed by  taking into account the geodesic potential $V_{\rm gK}(r)$.
We have derived the potential $V_i(r)$ when expanding $V_{\rm K}(r)$ at large distance $r$ because  $V_{\rm K}(r)$ is a complicated form to analyze superradiant instability.
Among $V_i(r)$, two lowest order potentials [$V_1(r)$ and $V_2(r)$] are suitable for analyzing the superradiant instability.
$V_1(r)$ is the Newtonian (attractive) potential, while $V_2(r)$ includes centrifugal (repulsive) term if Eq.(\ref{l-bound}) is satisfied. We note that  $V_2(r)$ matches  $V_{\rm gK}(r)$ up to  $(1/r)^2$-order, but $V_3(r)\sim V_{\rm gK}(r)$.

It is clear that all conditions for superradiant instability include $\omega<\omega_c$  and  $\omega<\mu$  together with  a positive trapping well.
One condition ($V'_1(r)>0 \to M\mu^2>0$ ) for a trapping well is always satisfied for a massive scalar propagating on the Kerr spacetime because the Newtonian potential [$V_1(r)]$ is attractive.  Moreover, this condition implies either a trapping well or a tiny well. Hence, we need to seek for the other condition to get a trapping well.
The other condition could be found  by  solving the Schr\"{o}dinger equation together with $V_2(r)$ in the far-region.
The other condition takes the lower bound of  $a<0$ in $U(a,b;cr)~(\to M\mu^2/\sqrt{\mu^2-\omega^2}>k+1/2)$ to obtain quasibound states.
This condition may describe that the attractive force is greater than the repulsive force.
In  case of $a>0(\to M\mu^2/\sqrt{\mu^2-\omega^2}<k+1/2)$, we have a tiny well located very far from the outer horizon to get bound states.   This condition implies that the attractive force is less than repulsive force. The case of $a\simeq0$ indicates the boundary between  trapping well and tiny well.

Finally,  it is interesting to note that an ultralight boson has a tiny well located at the asymptotic region ($r=8.0\times 10^4$), possessing a bound state.
If superradiant instabilities are ubiquitous to all Kerr black holes for a massive scalar satisfying $\omega<\omega_c$ and $\omega<\mu$~\cite{Ghosh:2021uqw}, we have to obtain superradiant instability for two potentials with a tiny well [(Left) Fig. 2 and (Left) Fig. 5] with $M\mu\sim (0.01,0.1)$.  However, our potential analysis shows superradiant stability because these have still asymptotic bound states [see Fig. 14]. In this case, we  may solve Eq.(\ref{wave-rad}) directly to find out $\omega_{\rm R}$ and $\omega_{\rm I}$~\cite{Detweiler:1980uk}. We suggest that a tiny well plays an important role of  making $\omega_{\rm I}>0$ to give superradiant instability. In this case, one has to explore a connection between tiny well and $\omega_{\rm I}>0$.

 \vspace{0.5cm}

{\bf Acknowledgments}
 \vspace{0.5cm}

This work was supported by a grant from Inje University for the Research in 2021 (20210040).

\end{document}